# Constraint Exploration and Envelope of Simulation Trajectories

Oswaldo Terán[1], Bruce Edmonds and Steve Wallis
{o.teran, b.edmonds, s.wallis}@mmu.ac.uk
Centre for Policy Modelling
Manchester Metropolitan University
Aytoun Building, Aytoun Street, Manchester, M1 3GH, UK.
http://www.cpm.mmu.ac.uk/

**Abstract**

The implicit *theory* that a simulation represents is precisely *not* in the individual choices but rather in the 'envelope' of possible trajectories – what is important is the shape of the whole envelope. Typically a huge amount of computation is required when experimenting with factors bearing on the dynamics of a simulation to tease out what affects the shape of this envelope. In this paper we present a methodology aimed at systematically exploring this envelope. We propose a method for searching for tendencies and proving their *necessity* relative to a range of parameterisations of the model and agents' choices, and to the logic of the simulation language. *The exploration consists of a forward chaining generation of the trajectories associated to and constrained by such a range of parameterisations and choices.* Additionally, we propose a computational procedure that helps implement this exploration by translating a Multi Agent System simulation into a constraint-based search over possible trajectories by 'compiling' the simulation rules into a more specific form, namely by partitioning the simulation rules using appropriate modularity in the simulation. An example of this procedure is exhibited.

KEYWORDS: Constraint Search, Constraint Logic Programming, Proof, Emergence, Tendencies

## 1. Constraint Declarative Programming and Exploration of Simulation Trajectories

Several approaches have turned up as an answer to the need for declarative programming with a more flexible manipulation of the semantic than traditional Logic Programming (LP) and forward chaining systems do. The call has been for techniques allowing a *semantic driven search* (Frühwirth et al., 1992).

The first answer came from Logic Programming, *Constraint Logic Programming*, commonly using Prolog, both as a platform and as the programming style (Frühwirth et al., 1992). It is based in backward chaining inference. A second answer came from *Rule Based Forward Chaining systems*. Examples are Constraint Handling Rules (CHR and its improved version $CHR^v$; see Abdennadher et al., 1999) Constraint Rule Base Programming (CRP) (Liu et al.), Satchmo and CPUHR-tableux calculus (Addennadher, 1995; Idem, 1997). Among the advantages of these over CLP are allowing alternative logical extensions via split and backtracking (e. g. Satchmo, $CHR^v$, CRP), introduction of user defined constraints (e.g. CHR) and Meta and Higher-Order reasoning via re-writing of rules (e. g. Satchmo).

---

[1] Also a member of the Centro de Simulación y Modelos (CESIMO: Centre for Simulation and Modelling) and the Department of Operations Research of the University of Los Andes, Venezuela (http://cesimo.ing.ula.ve/).

Until now Constraint Logic Programming's aim has been to search for a solution or satisfaction of a goal. In CLP the aim is to look for a proof (like in LP) while in constraint forward chaining systems to find a model satisfying certain conditions has been the purpose. So, in the first approach the conclusion is based (in some sense) in a whole exploration while in the second only one among the possible solutions is searched (one extension).

A similar situation to this in LP has appeared in other areas of research, for example, in simulation – a model oriented search approach. There usually constraints are given via fixed parameters of the model and choices representing simulation optional trajectories. For example, in Multi-Agent Systems (MAS) choices might be due to agents' decisions.

In traditional simulation, the dynamics or the program is generally analysed using *Scenario Analysis* (Domingo et al., 1996) and *Monte Carlo* techniques (Zeigler, 1976). In the first the dynamics of each trajectory is considered while in the second a sample of them is studied through statistical techniques. Due to the fact that a lot of computational resources are required in the former and several not always desirable assumptions and simplifications are made in the later, there is still a need for more appropriate techniques for analysing simulation dynamics. For example, in social simulation it is important to search for tendencies being *non-contingent*, that is, being common to several paths, and *non-expected* from the simulation design (Edmonds, 1999). These are called *emergent* tendencies. The theoretical value of the analysis of the dynamics of a simulation can be seen, for example, in *alignment* of models (Axtell et al., 1996).

On the other hand, *declarative programming* seems suitable for simulation due to its flexibility, expressiveness and the correspondence of the simulation extension to certain logic extensions allowing formalisations and a promise for formal proofs. In this sense, declarative programs open new ways for exploring simulation dynamics, apart from the named traditional methods. However, in applications like social simulation proving usually is almost impossible because of the huge amount of required computation – too many simulation trajectories appear.

*Constraint logic programming* seems promising for alleviating these drawbacks. A systematic - controlled constraint search for alternative trajectories in a simulation will allow bring in stronger conclusions and more assertive and fruitful theoretical exploitation of the experience than when using traditional tools. Constraint search is not something new in declarative programming. Both backward and forward chaining inference based tools have been developed for constraint reasoning. However, these approaches could not be straightforwardly appropriated for simulation applications.

Our main *goal* in this presentation is to propose *a framework for a constraint search of tendencies* in simulation trajectories and a technique for implementing it in a declarative simulation language. Constraints in simulation are due to parameters of the model and choices, where each 'choice' means that the simulation takes one of the possible 'trajectories'. We are particularly interested in searching for *emergent* tendencies. The search will be in a subspace of trajectories defined by the range of allowed constraints and parameters, and the logic of the simulation language. Additional advantages to make the exploration efficient can be taken from this semantic driven search -it will be the case in the technique to be proposed. We used a declarative simulation language (SDML: Strictly Declarative Modelling Language; Moss et al., 1998) suited for constraint searches because of its facilities for backward and forward simulation, backtracking, re-writing of rules and an internal assumption manager allowing certain predefined manipulations of constraints.

We will begin in section 2, by outlining the main features of SDML for a constraint search. The implementational concerns of the technique, i.e. the proposed architecture for doing the constraint-based model search in a "hunt" of tendencies is described in section 3. Following this (section 4), we will give an example of a technique applying this architecture. Then in section 5, we will present a case where this methodology using the technique previously described is used. In section 6, we briefly position this approach with respect to general theorem proving, proving in Multi-Agent Systems and constraint logic programming. Finally, some conclusions are made.

## 2. Towards the implementation of a suitable platform for a constraint envelope of trajectories using SDML

SDML (Strictly Declarative Modelling Language; Moss et. al, 1998) is the declarative Multi-Agent System in which we have developed our experiments. As a source of comparisons and ideas, we have also programmed our model in the Theorem Prover OTTER (McCune, 1995; Wos et. al, 1965; Wos, 1988; Chiang et al., 1973).

### 2.1 Relevant characteristics and features SDML offers are
- Good underlying logical properties of the system. SDML's underlying logic is the Strongly Grounded Autoepistemic Logic (SGAL) described by Kurt Konolige (Konolige, 1995).
- Its backtracking procedure facilitates the exploration of alternative trajectories via the splitting of simulation paths according to agents' choices and model's parameters.
- The assumptions manager in SDML tracks the use of assumptions. Assumptions result from choices.
- A good collection of useful primitives relevant to, for example, social simulation.
- The type meta-agent. A meta-agent (meta, for our purposes) is an agent "attached" to another agent as a controller; it is able to program that agent. This is used here *not* as an agent *per se* but as a module used to 'compile' rules into an efficient form as well as to monitor and control the overall search process and goals.

### 2.2 Internal Manipulation of Constraints in SDML
A partition is a grouping of rules according to their dependencies SDML does. Dependencies among rules in different partitions give dependencies among partitions. Rules in a partition do not have dependencies on the subsequent partitions. Assumptions are made for each partition in accordance to choices in such a partition.

The SDML assumption manipulator is a sort of Truth Maintenance System (TMS) for each partition (see figure 1). For example, when certain value for a variable has been deduced under two different assumptions, then a disjunction of the two original assumptions is placed for this datum in the database. This reasoning is helped by the introduction of choices, when defining data and applying rules, and backtracking. Once a

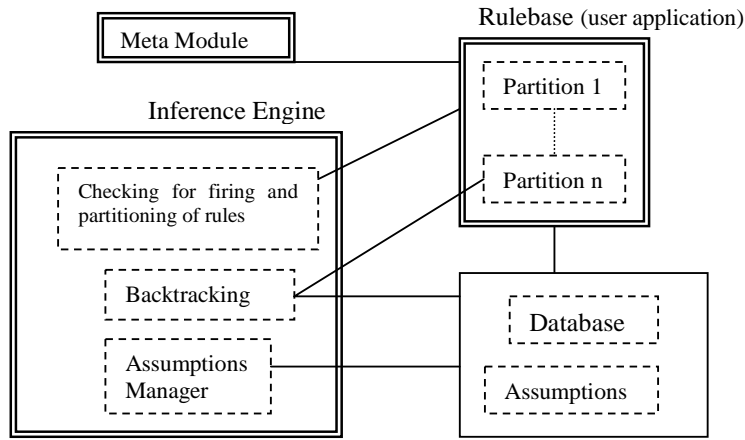

**Figure 1.** Overview of SDML's framework.

contradiction (e.g. the predicate false in the consequent of a rule) is found the system backtracks and a new choice is made in that partition. When all choices have been unsuccessfully tested in a partition the system backtracks to a previous partition to make a different choice there.1

Meta is a module to write rules in the system at the beginning of the simulation. It permits manipulations and reasoning in a higher level. Its semantic manipulation obviously increases flexibility, e.g. to write rules referring explicitly to instances of a type rather than to types. Meta will be one of the corners stones of the technique to be presented. Such a manipulation will be very useful both, for driving the search and for making it efficient in terms of computational time.

In SDML, choices are introduced via *built in* predicates. For example, the primitive *randomChoice* allows choosing randomly from several alternative values. Each choice will define a different simulation path labelled with an assumption. Another example is *notInferred*, one of the primitives allowing non-monotonic reasoning, which allows generation of data when a certain fact is not present in the database, e.g. *notInferred b, implies c*, will put c in the database if *b* is not in the database. If this value, *b*, is written later during the search in the database, then the assumption becomes false. In case the rule written *b* is in a different partition from the one where the assumption was made, then *c* and any other data supported by the assumption is withdrawn via backtracking to the partition where the assumption was made.

## 3. Constraint Model Generation and Envelope of Trajectories

### 3.1 Constrained exploration of trajectories

We propose the use of an exhaustive constraint-based search over a range of possible trajectories in order to establish the necessity of postulated emergent tendencies. Thus a subset of the possible simulation parameterisations and agent choices is specified; the target emergent tendencies are introduced in the form of negative constraints; and an automatic search over the possible trajectories is performed. (See figure 2).

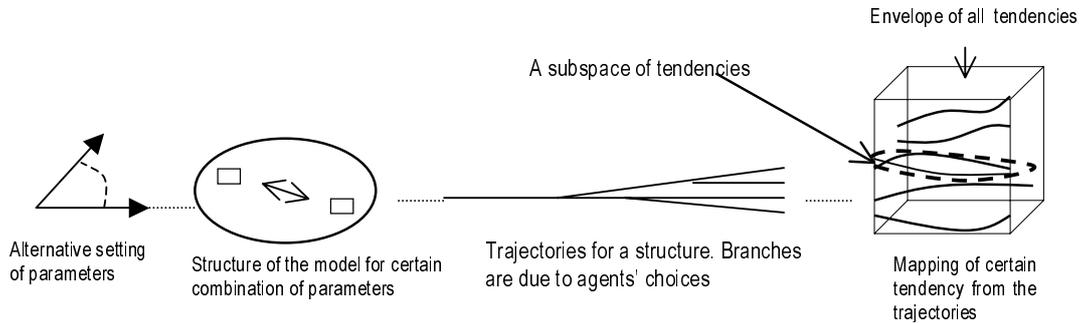

**Figure 2.** A constraint-based exploration of possible simulation dynamics

### 3.2 Characterising the envelope of tendencies

In order to distinguish between the *exceptional* and the *representative* in a simulation, we will formally describe the envelope of certain tendencies in a simulation. This might be done by:

- Certain *properties* satisfied by the observed tendency.
- A *mathematical description* of a subspace of the tendencies or of a subspace given a bound of the tendencies.
- *Representative or typical instances* of such a tendency.
- A *mapping* from the setting of trajectories, as given by the alternative arrangement of parameters and agents' choices, to certain knowledge (maybe properties) about the tendency: *(parameters* X *choices)*→ *(know. of the tend.)*

### 3.3 Proving the necessity of a tendency

We want to be able to *generalise* about tendencies going from observation of individual trajectories to observation of a group of trajectories generated for certain parameters and choices. Actually, we want to know if a particular tendency is a necessary consequence of the system or a contingent one. For doing this we propose to *translate* the original MAS along with the range of parameterisations and agents' choices into a platform (described in the next section) where the alternative trajectories can be *unfolded*. Each trajectory will correspond to a possible trajectory in the original MAS. Once one trajectory is shown to satisfy the postulated tendency another set of parameters and agents' choices is selected and the new trajectory is similarly checked. If all possible trajectories are successfully *tested*, the tendency is *proved to be necessary* relative to the logic of the simulation language, the range of parameterisations and agents' choices.

The idea is to translate the MAS into a constraint-based platform in an automatic or near automatic way without changing the *meaning* of the rules that make it up in order to perform this automatic testing. In this way a user can program the system using the agent-based paradigm with all its advantages; inspect single runs of the system to gain an intuitive understanding of the system and then check the generality of this understanding for fragments of the system via this translation into a constraint-based architecture.

In the *example* shown below, all trajectories are explored for one combination of parameters, eight agents' choices per iteration and seven iterations. A simple tendency was observed characterised by a mathematical description of its boundaries. This characterisation was handled as a theorem. The theorem was proved to be necessary following a procedure similar to the one described in the previous paragraphs.

### 3.4 What is new in this model-constrained methodological approach

It is our goal in this paper to propose an alternative approach for exploring and analysing simulation trajectories. It will allow the entire exploration and subsequent analysis of a subspace of the whole space of simulation trajectories. We are suggesting the generation of trajectories in a semantically constrained way. Constrictions will be context-dependent (over the semantics of the trajectory itself) and will be driven via the introduction of a controller or meta-module.

## 4. Implementing a suitable constraint-based programming platform for the envelope of trajectories

The main goal of the programming strategy to be described is to increase the efficiency in terms of simulation time, thus making an efficient constraint-based search possible. The improvements will be achieved by making the rules and states more context-specific. This enables the language's inference engine to exploit more information about the logical dependencies between rules and thus increase the efficiency.

### 4.1 An Overview of the system.

We implemented the proposed architecture in three modules; let us call them **model**, **prover** and **meta**. The following diagram illustrates this:

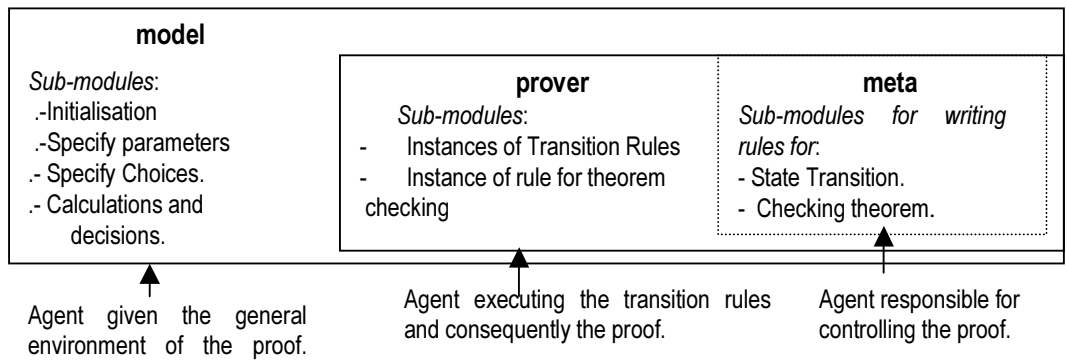

**Figure 3.** Physical view of the system.

### 4.2 Description of System Modules

We have found it convenient to *distinguish* and model as distinct entities three basic elements of a simulation: the *static structure* of the model, the *dynamics* of the simulation and the way this dynamics is "managed" by certain meta-rules or by a *controller*. Each of those entities is programmed in a different module:
1. **model**, *sets up the structure* of the model, that is, it gives the environment of the simulation: range of parameters, initialisations, alternative choices and basic (backward chaining) rules for calculations.
2. **prover**, *generates the dynamics* of the simulation. This is a sub-module of *model* (i.e. it is contained in *model*). This will basically contain the transition rules, auxiliary rules for generating pre-processing required data and the conditions to test the necessity of the theorem. All of them are rules to be executed while the simulation is going on.

3. **meta**, *is responsible for controlling the dynamics* of the simulation. Its meta-rules write the transition rules and the theorem in (as well as others required by) the module *prover*. A picture of the system is given in *Figure 3*.

**4.3 Program dynamics**

Modules' rules are executed in the following *sequence:*
1. **model**: initialising the environment for the proof (setting parameters, etc..)
2. **meta**: creating and placing the transition rules in *prover*.
3. **prover**: carrying on the simulation using the transition rules and backtracking when a contradiction is found.

The program *backtracks* from a path once the conditions for the theorem are verified, then a new path with different choices and/or parameters is picked up. Next figure describes a transition step.

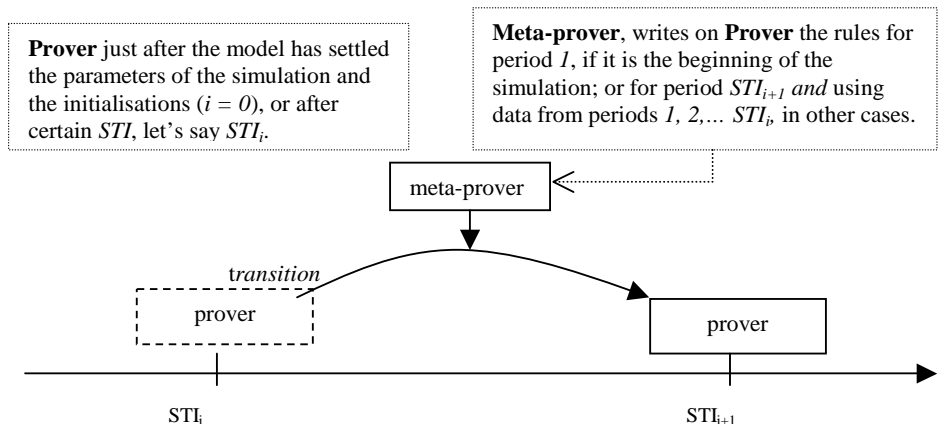

**Figure 4** State transition from $STI_i$ to $STI_{i+1}$

**4.4 Split of the rules: a source of efficiency.**

In forward chaining simulation the antecedent retrieves instance data from the past in order to generate data for the present (and maybe the future):

$$\text{past facts} \rightarrow \text{present and future facts}$$

Traditionally, the set of transition rules are implemented to be general for the whole simulation. A unique set of transition rules is used at any STI (Simulation Time Instant or iteration).

As the simulation evolves, the size of the database increases and the antecedents have to discriminate among a growing amount of data. At $STI_i$, there would be data from *(i-1)* alternative days matching the antecedent. As the simulation evolves it becomes slower because of the discrimination the program has to carry out among this (linearly) growing amount of data.

Using the proposed technique, we would write a transition rule for each simulation time. The specific data in the antecedent as well as in the consequent could be instanced. Where possible, a rule for each datum, the original rule will generate, would be written. This will be illustrated in the example of the next section.

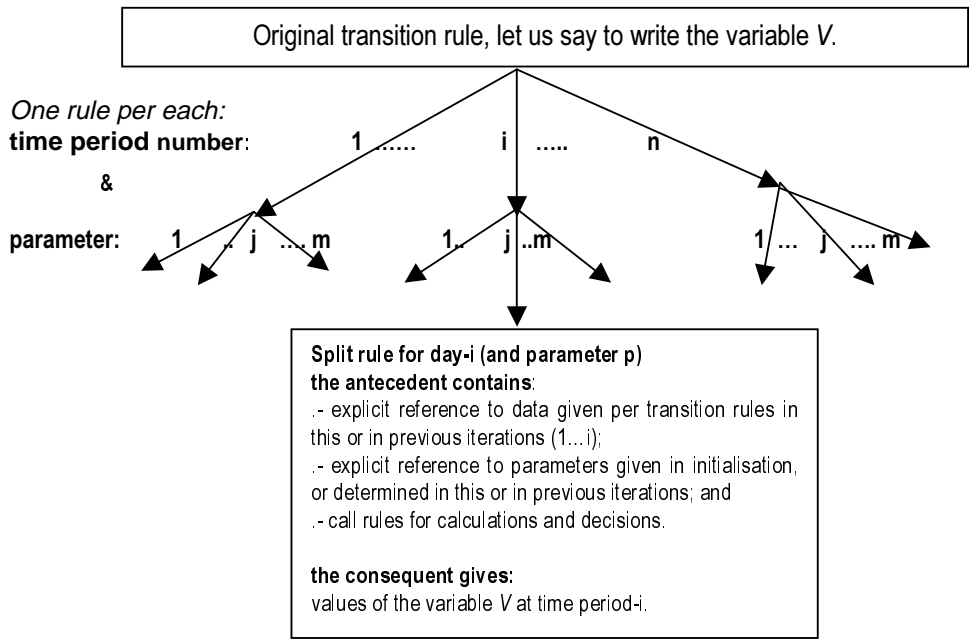

**Figure 5.** Splitting of rules by time period and a combination of parameters.

This technique represents a step forward in improving the efficiency of declarative programs. One could, in addition, make use of partitions and time levels to introduce further modularity – this would further speed up the search process and increase the memory that is needed. Partitions permit the system to order the rules to fire in an efficient way according to their dependencies. Time levels let us discriminate among data lasting different amounts of time. The *splitting of rules lets us discriminate among the transition rules for different simulation times given a more specific instancing of data at any STI.*

**4.5  Measuring the efficiency of the technique**
Comparing the two programs, the original MAS simulation and the constraint-based translation we obtain a *speed up* by a factor of $O(NM)$, where $N$ is the number of agents and $M$ is the number of STIs. SDML already has facilities for discriminating among STIs, but their use is not convenient for the sort of simulation we are doing (exploring scenarios and/or proving) because of the difficulties for accessing data from any time step at any time. If we had used this facility still the simulation would have been speeded up by $N$. Notice that all these values are only estimations because a program stops trying to fire a rule as soon as it finds out that one of its clauses is false.

It is clear that the greater the number of entities in the simulation or the number of STIs, the larger the benefits from the technique. We must notice that the speeding up of the simulation is only one dimension of the efficiency given by the technique.

**4.6  Translating a traditional MAS architecture into a model-exploration MAS architecture.**
Before splitting the rules the original MAS is reduced in a sort of *unencapsulation* of the hierarchy of agents into the architecture shown in figure 3. Additional variables must be added into predicates and functions in order to keep explicit the reference to the "owner" agent of any instance of a variable. This will facilitate the check for tendencies, the testing of the theorem and any other data manipulation. It is as if the agent where replaced by its rulebase, see figure 6.

In the original architecture, each agent has its own rulebase (RB) and database (DB). The agent's structure is given by its set <RB, DB> as well as by the structure of any subagents.

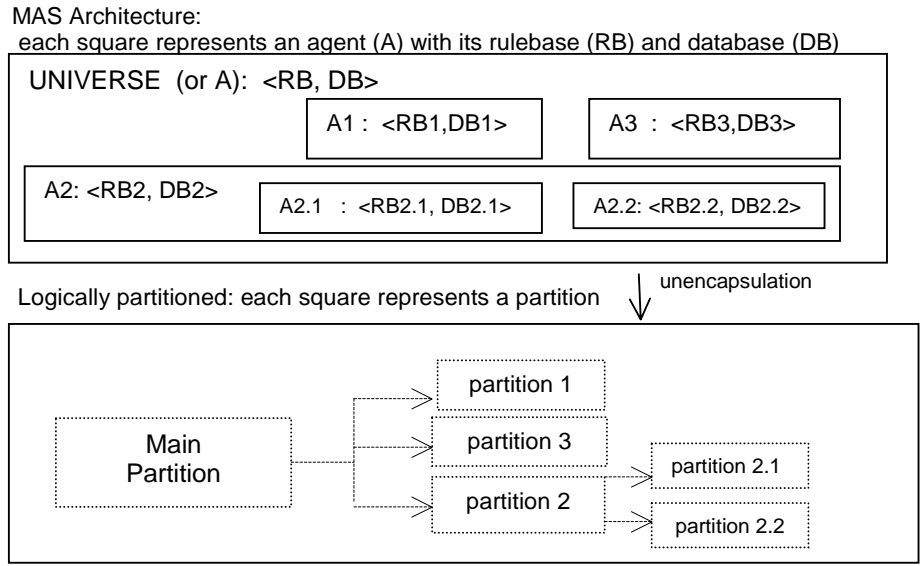

**Figure 6.** Unencapsulating a MAS' architecture

Using the technique, the initialisation of the static structure is accomplished by the module model, as explained above. The transition rules (dynamic structure) will be set up by the module meta into the module prover. There is still a hierarchy, both in the structure of the model and in the dynamics of the simulation – it is given by the precedence in the rulebase partition. Additional partitions will appear as the rules are split, discriminating among rules according to the factors used for splitting (e.g. STIs, consumers and producers in the example, see below).

## 5. An example.

A simple model of a producer-consumer system, which was previously built in SDML and in the Theorem Prover OTTER, was rebuilt using the proposed modelling strategy. In the new model the exploration of possibilities is speeded up by a factor of 14. Also, the model built in OTTER, though faster than the original model in SDML, is several times slower than the improved model built in SDML.

Some of the split transition rules were the ones for creating (at each STI) producers' prices and sales, consumers' demand and order, warehouses' level and factories' production. Among the rules for auxiliary data split were the ones for calculating: total-order and total-sales (a sum of the orders for all producers), total-order and total-sales per producer, and total-order and total-sales per consumer.

### 5.1 Example of a split rule: *Rule for prices*.

This rule calculates a new price for each producer at each STI (which we called *day*), according to its own price and sales, and the price and sales of a chosen producer, at the immediately previous STI.

The original rule in SDML was like this:
```
for all (producer)
for all (consumer)
for all (day)
(
      price(producer,myPrice,day)                     and
      totalSales(totalSales,day)                      and
      sales(producer,mySales,day)                     and
      choiceAnotherProducer(anotherProducer)          and
      price(anotherProducer,otherPrice, day)          and
      calculateNewPrice(mySales,totalSales,
      otherPrice,myPrice,newPrice)
   implies
      price(producer, newPrice, day + 1)
)
```

The new rule (in the efficient program) will be "broken" making explicit the values of prices and sales per each day.

In the following, we show the rule per *day-i* and *producer-j*:
```
for all (consumer)
(
      price(producer-j, myPrice, day-i)               and
      totalSales(totalSales, day-i)                   and
      sales(producer, mySales, day-i)                 and
      choiceAnotherProducer(anotherProducer)          and
      price(anotherProducer, otherPrice, day-i)       and
      calculateNewPrice(mySales,totalSales,otherPrice,
      myPrice,newPrice)
   implies
      price(producer-j, newPrice, (day-i) + 1)
)
```

If the name for the clauses *price* and *sales* (e. g.those clauses associated with the data recalculated at each STI) are used to make explicit the day, the rule will have the following form. It is important to observe that *only one instance of newprice in the consequent is associated with only one transition rule and vice verse*:
```
for all (consumer)
(
      price-i(producer-j, myPrice)                    and
      totalSales-i(totalSales)                        and
      sales-i(producer-j, mySales)                    and
      choiceAnotherProducer(anotherProducer)          and
      price-i(anotherProducer, otherPrice)            and
      calculateNewPrice(mySales,totalSales, otherPrice,
      myPrice,newPrice)
implies
      price-(i+1)(producer-j, newPrice)
)
```

**5.2 Other facts about the example**

There were eight "types" of rules (see table below) involved in the application of the technique, those associated with generating the dynamics of the simulation. Excluding the rules for testing the theorem and setting up producers' choices, which did not suffer additional split, the other six rules in the original model where split into 96 rules in the new model. All of them were split by transition time step (six transitions); among these six, two suffered additional split by producer (there were three producers), and one, among the last two, was also split by consumer (there were three consumers). This gives: *(4 + (1 + (1 * 3)) * 3) * 6 = 96* rules in the new model replacing the referred six rules in the old model.

| Description of the rule (rule for:) | No. of rules in the original model | Rule split by | No. of rules in the new model |
|---|---|---|---|
| Checking theorem | 1 | --- | 1 |
| Producers' choice of another consumer for comparing certain data when changing price | 6 | already split by STI | 6 |
| Calculating total orders by cons. and prod. | 1 | STI | 6 |
| Calculating total producers' sales | 1 | STI | 6 |
| Determining (D) consumers' demand and order | 1 | STI | 6 |
| D. Producers' price | 1 | STI | 6 |
| D. Producers' production and level in store | 1 | STI and producer | 18 |
| D. Producers' sale | 1 | STI, producer and consumer | 54 |

**Table 1.** Comparing the number of rules in the original and in the new implementation.

Though splitting of the rules increases the necessary amount of memory for keeping the rules, it together with partitioning of the rulebase according to rule dependencies, which has already been implemented in SDML, allows a much faster instantiation of the rules and consequently speeds up the simulation. This is due to the fact that rules refer to data more explicitly so that when intending to fire a rule the data considered to check the antecedent of the rule is limited to a smaller part of the database. The size of the searched part of the database keeps stable as the simulation time progresses while without splitting this search space, in SDML, might grow linearly as the simulation time goes forward. Rules split by only STI might also have been "broken" by consumer and producer, but it was not necessary as a model fast enough for our purpose of proving the necessity of certain tendency was obtained with the described implementation. It is more in our interest to use this experience more intensively in future modelling.

**5.3 What the technique enables**

In this example, the described technique was used to prove that the size of the interval of prices (that is*: biggest price - smaller price,* each day) decreases over time during the first six STIs over a range of one parameterisation and eight choices for the agents at each STI. An exponential decrease of this interval was demonstrated in all the simulation paths. A total of 32768 simulation trajectories were tested. It was not possible to simulate beyond this number of days because of the limitations imposed by computer memory. The complete search process took only 24 hours.

Though the tendency we have shown is simple and quantitative, it is obvious that the technique is applicable in more interesting cases of emergent tendencies, even if they have a qualitative nature.

This technique is useful not only because of the speeding up of the simulation but also for its appropriateness when capturing and proving tendencies under the specified constraints. In the example, the meta-module was used to write the rule with the hypothesis (theorem) to be tested on prover-module at the beginning of the simulation. If the meta-module were able to write rules on prover-module while the simulation is going on, the theorem we wanted to prove could be *adapted* according to the results of the simulation via *relaxing constraints*. For example, the technique could be implemented in a way that we only give the program hints related to the sort of proof we are interested in. Then the meta-module would "*elaborate"*, via adapting over time in a context dependent manner, a set of hypotheses or theorems.

## 6. Other Approaches

### 6.1 Using OTTER (McCune 1995), a resolution-style first order Theorem Prover

In simulation, strategies like a Future Event List (FEL), in event-driven simulation, and partition of the space of rules and a hierarchy of nested time levels, in declarative simulation systems (e.g. some MAS), are used. The criteria for firing rules is well understood, and procedures like weighting and subsumption usually are not necessary. Additionally, redundant data for some purpose could be avoided in MAS with appropriate compilation techniques.

The advantages given for the weighting procedure in OTTER are yielded in MAS systems like SDML by procedures such as *partitioning*, where chaining of the rules allows firing the rules in an efficient order according to their dependences.

The main difficulty for our simulation purposes when using OTTER was the lack of facilities for accessing the database. Clearly the introduction of a meta-module and other facilities for reasoning about predicates and rules in SDML brings a great improvement in this sense.

### 6.2 DESIRE

Among other approaches for the practical proof of MAS properties, the more pertinent might be the case conducted by people working in DESIRE (Engelfriet et. al., 1998). They propose the hierarchical verification of MAS properties, and succeeded in doing this for a system. However, their aim is the verification of a computational program – it is proved that the program behaves in the intended way. It does not include the more difficult task, which we try to address, of proving properties of the simulation dynamics.

### 6.3 Satchmo and other CLP constraint programs

Some of Satchmo's and other constraint programming languages' facilities are similar to SDML's ones, for example, *backtracking* and Satchmo's *false* predicate. However, they present certain built-in facilities for manipulation of constraints that SDML has not. For instance, reasoning about terms in CLP(X) or consistence techniques to prune the range of trajectories in other CLP (Frühwirth, 1992). Instead, SDML allows facilities to introduce alternative values for the manipulated entities (e.g. predicates, clauses, integer variables) which can be used as *constraints* (clauses for choosing, e.g. randomChoice) as well as a *meta module* able to reason about terms or rules. A meta module can build rules taking advantage of the simulation semantics. Permitting it to act while the simulation is going on will allow it to adapt the search to the simulation results. Because of all this SDML is able to control the manipulation of constraints flexibly and transparently for the user.

On the other hand, in CLP the aim is to look for a proof while in constraint forward chaining systems the purpose is to find a model satisfying certain conditions. In the first group of languages the conclusion is based (in some sense) in a whole (implicit) exploration of trajectories while in the second group only one among the possible solutions is searched (based on one trajectory). In contrast we have proposed a methodology for searching and proving the necessity of a tendency in a subspace of trajectories, which seems more appropriate in many applications where a broaden proving is prohibitive due to the huge amount of required computational resources.

It should be possible to rewrite the example in some of these programs, for instance in CHR$^v$. It is our purpose to enrich our methodological approach from the experience of people working in constraint programming. It will be one of our tasks to rewrite future models in some of these languages in order to bring in new ideas.

## 7. Conclusions and future work

We have argued and shown using an example the pertinence of a methodology for a constrained exploration and envelope of trajectories as a complement to traditional methods dealing with post-hoc analysis of the dynamics of simulations. We have suggested a forward chaining semantically constrained generation of trajectories.

Like in Constraint Logic Programming the constraint of the generation of simulation extensions respond to the need for a systematic and controlled exploration of the simulation dynamics. Our aim is to analyse subspaces of trajectories rather than searching for a model. When searching for a model, a "committed" or "non-determinist" (do not care about determinism in Abdennadher words; Abdennadher, 1999) search, i.e. search for an extension without backtracking, is done.

A platform to implement this methodology has been proposed. It consists of a modular structure according to strategic parts of a simulation: a first module, *model*, sets up the *static structure* of the simulation; then a second module, *prover*, generates the *dynamics* of the simulation; and finally a *meta-module* is responsible for *controlling* the dynamics of the simulation. The second characteristic of this platform is a *partitioning* of the space of rules and *splitting of transition rules* by STI, parameters and choices.

The control of the search via a *meta-module* makes the manipulation of the constraints more flexible, transparent and handy for the user (a controller in an even higher level) than when it is introduced into the same level than the transition rules of the simulation. Constrains are context-dependent (over the semantic of the trajectory itself) as the meta-module is *able to access the semantics* of the simulation setting up in advance one among the possible combination of agents' choices and model parameters for each simulation run.

In the offered example only a combination of parameters but all possible combinations of agents' choices were followed. The meta-module sat up a possible combination of choices after the combination of parameters of the model was defined and changed it after backtracking. Backtracking occurred as soon as certain conditions characterizing certain tendency where identified in the trajectory. This allowed prove the necessity of the tendency for that combination of parameters and the range of agents' choices. Thought in the example the tendency was identified by the modeller, also a monitoring module aiming to recognize relevant tendencies might have been introduced.

In this example the meta-module acts only at the beginning of the simulation. More powerful and flexible programs are possible when the meta-module is able to act at any STI. Transition rules can be evolving, in a sense that the meta-module builds the rules for a STI once every fact for the previous STI are known. This allows adaptation (e.g. relaxing) of the constraints defining the theorem according to the "discovered" circumstances into the explored dynamics of the simulation.

The *splitting* of the transition rules represents a step forward in improving the efficiency of declarative programs, one additional to the use of partitions and time levels. It lets us *discriminate* among the transition rules for different simulation time steps given a more *specific* instancing of data at any time step. Thus this alleviates some of the drawbacks of declarative programming due to the necessary grasping and updating of all state variables at any STI.

Our future work will be focused both in the search for a refining and improving of the presented platform as well as in its application for proving more interesting cases of emergent tendencies.

## Acknowledgements

SDML has been developed in VisualWorks 2.5.2, the Smalltalk-80 environment produced by ObjectShare. Free distribution of SDML for use in academic research is made possible by the sponsorship of ObjectShare (UK) Ltd. The research reported here was funded by CONICIT (the Venezuelan Governmental Organisation for promoting Science), by the University of Los Andes, and by the Faculty of Management and Business, Manchester Metropolitan University.